\documentclass[floatfix]{revtex4}

\usepackage{amsfonts,amsmath,amssymb}
\usepackage{graphicx,color}
\usepackage{verbatim}

\begin{document}

\title{Simulation of the shape memory effect in a NiTi nano model system}

\author{Daniel Mutter}
\email{daniel.mutter@uni-konstanz.de}
\affiliation{Department of Physics, University of Konstanz, 78457 Konstanz, Germany}
\author{Peter Nielaba}
\affiliation{Department of Physics, University of Konstanz, 78457 Konstanz, Germany}

\begin{abstract}
The shape memory behavior of a NiTi nanoparticle is analyzed by molecular dynamics simulations. After a detailed description of the equilibrium structures of the used model potential, the multi variant martensitic ground state, which depends on the geometry of the particle, is discussed. Tensile load is applied, changing the variant configuration to a single domain state with a remanent strain after unloading. Heating the particle leads to a shape memory effect without a phase transition to the austenite, but by variant reorientation and twin boundary formation at a certain temperature. These processes are described by stress-strain and strain-temperature curves, together with a visualization of the microstructure of the nanoparticle. Results are presented for five different Ni concentrations in the vicinity of 50\%, showing for example, that small deviations from this ideal composition can influence the critical temperature for shape recovery significantly.
\end{abstract}

\maketitle

\section{Introduction}

Due to their remarkable mechanical properties upon loading followed by heating [``shape memory effect'' (SME)] and vice versa [``superelasticity'' (SE)], shape memory alloys (SMAs) are widely applied in these days in diverse macroscopic engineering devices \cite{buchlag:10}. Microscopically, these effects result from the martensitic phase transformation between crystal structures of different symmetries, and from the formation of differently oriented variants of the low-temperature ground state phase (``martensite''). The question, whether and how this behavior takes place in systems with very small lengths scales, where a strong influence of the free surfaces comes into play, is of increasing interest in research and technology, for example in order to construct functional nanoscale actuators \cite{bhajam:05,bhacor:07}.\\
A lot of effort has been done in studying the martensitic and reverse (``austenitic'') transformation in nanosized systems (an overview is given in \cite{waitsu:09}), where for example new types of domain configurations emerge upon cooling of nickel-titanium nanocrystals below strongly size dependent transition temperatures \cite{waiant:08}. Experimental studies of SE and SME of NiTi \cite{clagia:10} and CuAlNi nanopillars \cite{sanno:08} show, that these effects take place at least down to length scales of several hundred nanometers, but, in case of CuAlNi, there is a considerable size dependence of the stress-strain curves if compared to bulk systems \cite{sanno:09}.\\
Besides experiments, computer simulations provide a tool for investigating SME and SE at an atomic scale. By applying loading, unloading and heating conditions to a martensitic NiAl system under periodic boundary conditions, total deformation recovery was obtained in a molecular dynamics (MD) simulation \cite{uehtam:06}. Strong zigzag behavior of the stress-strain curve upon loading due to abrupt stress releases was shown to be smoothed by differently oriented martensitic grains as starting configuration \cite{uehasa:09}, which fits the realistic situation in a better way. Concerning nanosized systems, the austenite-martensite transformation upon loading, which is the origin of SE, has been studied by MD by straining a NiTi nanocrystal with free surfaces, leading to stress decrease if martensite nucleates heterogeneously in the sample \cite{satsai:06}. By applying MD, several characteristics of structural phase transformations in NiTi model systems have been investigated in the past by Suzuki and coworkers \cite{suzshi:01,suzshi:03,dendin:10} with a pair potential, as well as by Ishida and Hiwatari \cite{ishhiw:07}, who used the modified embedded atom method. A new type of SME in fcc metal nanowires, triggered by an extremely high surface-to-volume fraction and the one-dimensional character of the system, could be identified by MD \cite{pargal:05,liazho:06}. Here, a structure conserving change of the lattice orientation propagates through the wire in order to accommodate the external load along the wire axis, leading to strains of over 50\%. Heating above a critical temperature changes the configuration back to the initial lowest energy orientation due to strong surface stresses, whereupon the strain is totally released.\\
By performing MD, a related kind to this nanoscale shape memory behavior is studied in the present work in a NiTi model system with different Ni content. The term ``model system'' is used, since the applied interatomic potential leads to a lowest energy martensitic ground state structure, which has strong similarities to the experimentally observed monoclinic B19$'$, without being exactly the same. Nevertheless, the simulated nanocrystals contain of differently oriented variants in the martensite, what can be explained in general by minimization of the total energy, composed of surface, bulk and domain wall parts. An external load is applied to the system, leading to a reorientation of the variants at a critical stress, accompanied by a plateau in the stress-strain curves. A permanent strain remains after relaxing the load, but it is released upon heating above a strongly concentration dependent temperature. Since these transition temperatures, at which the original domain configuration is recovered, lie below the austenitic start temperatures, this kind of SME at the nanoscale is conspicuously different to the mechanism in conventional bulk SMAs.

\section{Methods and structural analysis}

\begin{figure}
\centering
\includegraphics[width=0.7\linewidth]{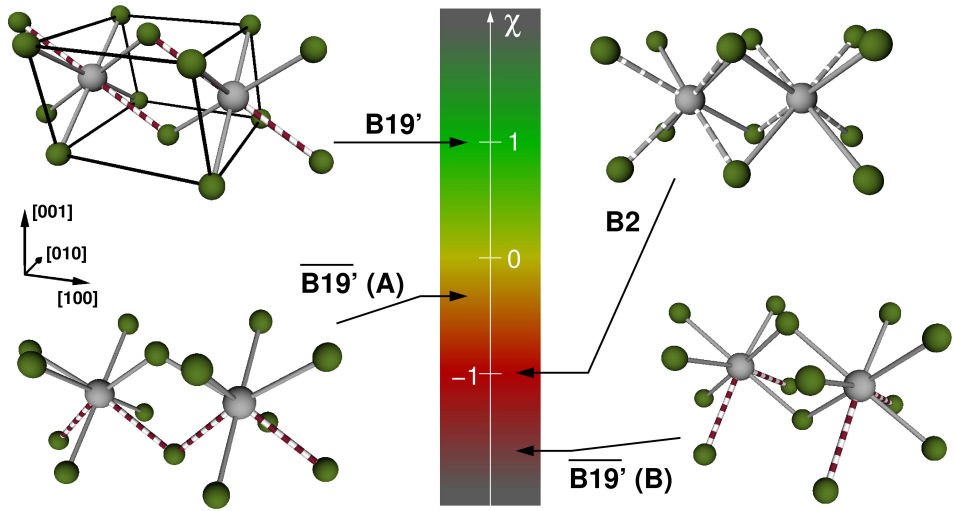}
\caption{Next neighbor environments of the structures and variants produced by the NiTi model potential. The dashed lines correspond to elongated lengths compared to the others [13.5\% in B19$'$ and $\overline{\mbox{B19}'}$ (A,B), 3\% in B2]. $\chi$ is an order parameter, with which the structures can be resolved on an atomic scale, and can be visualized by a color coding.}
\label{f1}
\end{figure}

In the the present study, a semiempirical interatomic potential of the Finnis-Sinclair type \cite{finsin:84} is used. It has its origin in the tight binding method in second moment approximation of the electronic density of states \cite{cleros:93}. The parameters for describing a NiTi alloy, i.e. values for Ni-Ni, Ti-Ti, and Ni-Ti interaction, have been determined by Lai and Liu \cite{lailiu:00}. The authors of the present paper have shown in a recent work \cite{mutnie:10}, that the application of a cut-off function proposed by Baskes \emph{et al.} \cite{basnel:89} with a width of $\delta$\,=\,$0.2\ \mathring{\mbox{A}}$ results in a stable monoclinic B19$'$ structure in a bulk system at low temperatures, with lattice parameters comparing well with experimental \cite{prokor:04} and \emph{ab initio} \cite{hatkon:09.2} values. In addition, an entropically stabilized nearly cubic B2 like structure is obtained upon heating. If the system is cooled out of this high-temperature phase, a martensitic ground state structure denoted $\overline{\mbox{B19}'}$ emerges. This structure is closely related to B19$'$, since it shows nearly the same peak structure in the radial distribution function, but the angular distribution of the nearest neighbor lengths is different, resulting in a ground state energy, which is about 0.04\% lower than that of B19$'$ \cite{mutnie:10}. In that work, periodic boundary conditions were applied in order to simulate bulk behavior. The present paper deals with a nano structure, where angular distributions of the nearest neighbor lengths in $\overline{\mbox{B19}'}$ are observed, which differ slightly from the one described in \cite{mutnie:10}. At free surfaces, these structures, denoted A and B, are energetically more favourable. Fig.\:\ref{f1} shows all these structures resulting from the model potential, together with an order parameter $\chi$, which is calculated out of the nearest neighbor environment of each atom \cite{mutnie:11}. Due to the shear in the monoclinic B19$'$ structure, two nearest neighbor lengths are elongated about 13.5\% compared to the other six, denoted by a dashed connection between the atoms. In the B2 like high temperature state, the lengths are all nearly the same, differing only about 3\% in another distribution. As mentioned above, in the $\overline{\mbox{B19}'}$ configuration, the elongations of nearest neighbor lengths are as long as in B19$'$, but they are arranged in another way. The order parameter $\chi$ is calculated out of the nearest neighbor distances, relative to an underlying B19$'$ reference structure with determined shear direction [100]. This leads to different values of $\chi$ in $\overline{\mbox{B19}'}$, depending on the direction of the elongated lengths relative to [100] ([100] A, or [010] B). These are just differently oriented variants of the same structure.

For symmetry reasons, one could think of a $\overline{\mbox{B19}'}$ version C as well, where the elongated neighbor lengths point into the [001] direction. In contrast to the versions A and B, we could not identify its existence in the martensitic state of the simulated nanoparticle. As explained below in more detail, the formation of domain walls between different orientations of the same structure costs a certain amount of energy. At small system sizes, this loss in energy is higher than the energy gain due to a possible free surface with a third orientation C, leading to only two observed variants in our system. Since the martensitic ground state of the used model potential is the $\overline{\mbox{B19}'}$ structure, which is not observed to exist experimentally in NiTi, the system is denoted as a NiTi model system in the following. In this sense, the present study deals with the general behavior of a nano shape memory alloy consisting of differently oriented variants, where load and temperature is applied, and explains it from a fundamental point of view, rather than reproducing exact NiTi behavior. In addition, it is investigated, how the behavior, which is mainly triggered by a huge surface-to-volume fraction, changes, if the concentration of a perfectly ordered bi-atomic alloy is altered slightly. The MD simulations are performed with a velocity-Verlet algorithm \cite{swoand:82} (timestep $10^{-15}$ s), and the temperature is controlled by a Nos\'e-Hoover thermostat \cite{nos:84}.

\section{Results and discussion}

\subsection{Martensitic variant configuration}

\begin{figure}
\centering
\includegraphics[width=0.7\linewidth]{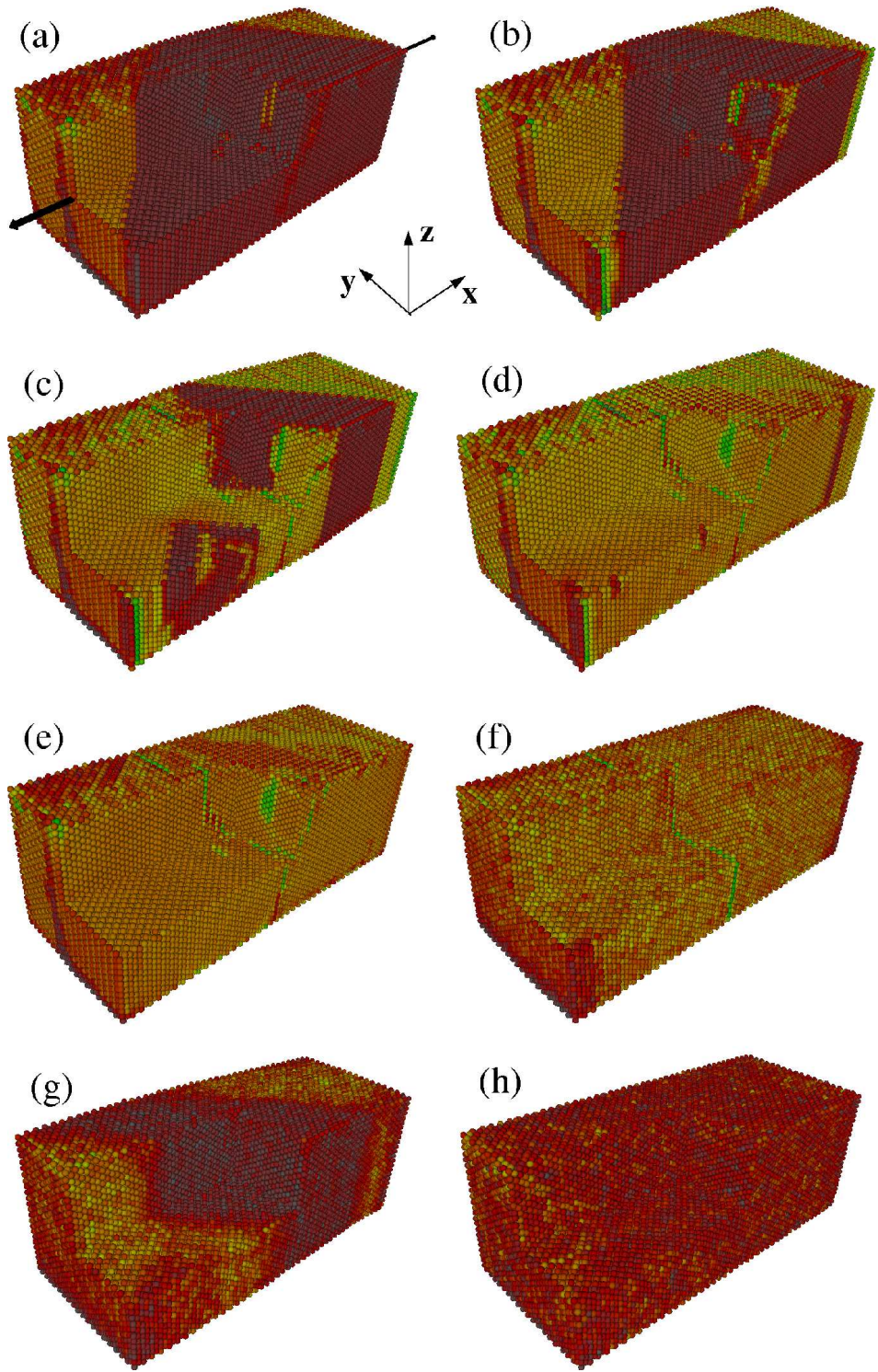}
\caption{Nanoparticle with dimensions 20x8x8 nm$^3$ in martensitic start configuration (a), during loading [(b)-(d)], unloading (e) and heating [(f)-(h)] process.}
\label{f2}
\end{figure}

In the following, a NiTi nanoparticle with dimensions 20x8x8 nm$^3$, consisting of 90000 atoms, is considered, with an ordered alloy structure and a concentration of nickel varying between 50\% and 52\%. In the martensitic state, which can be obtained by cooling a B2 system below the martensitic transition temperature, the particle consists of differently oriented $\overline{\mbox{B19}'}$ domains (twins), in a configuration as shown in Fig.\:\ref{f2}(a). This type of domain structure is determined by the geometry of the nanoparticle, since it results from an interplay between surface energy contributions and the energy of twin boundaries. At the free y-z-surface, the A orientation of $\overline{\mbox{B19}'}$ is energetically more favorable, while at the free x-z-surface it is the B version. This can be explained easily, since there is an angle of 90 degrees between the y-z- and the x-z-surfaces, and the $\overline{\mbox{B19}'}$ orientation B is obtained by rotating the version A by the same angle. So, both surfaces have the same atomic configuration. This is only possible, if there are twin boundaries in the system between variants A and B. Since the formation of twin boundaries costs an amount of energy $\Delta E_{tb}$, a multi variant martensitic structure in the nanoparticle only forms, if the energy gain due to the surface reorientation is bigger than $\Delta E_{tb}$. Then, the total energy of the multi variant configuration is lower than that of a single oriented $\overline{\mbox{B19}'}$ domain. The dimension of the variants depends on the relation between the surface areas, as was observed in the simulation of 40x5x4 nm$^3$ and 30x4x4 nm$^3$ systems, where much smaller $\overline{\mbox{B19}'}$ (A) domains were formed in the martensitic transition. This kind of geometry dependent variant configuration, triggered by the interplay between surface and twin boundary energies, might be a general attribute of nanoscale martensitic transforming systems.

\subsection{Tensile loading}

\begin{figure}
\centering
\includegraphics[width=0.7\linewidth]{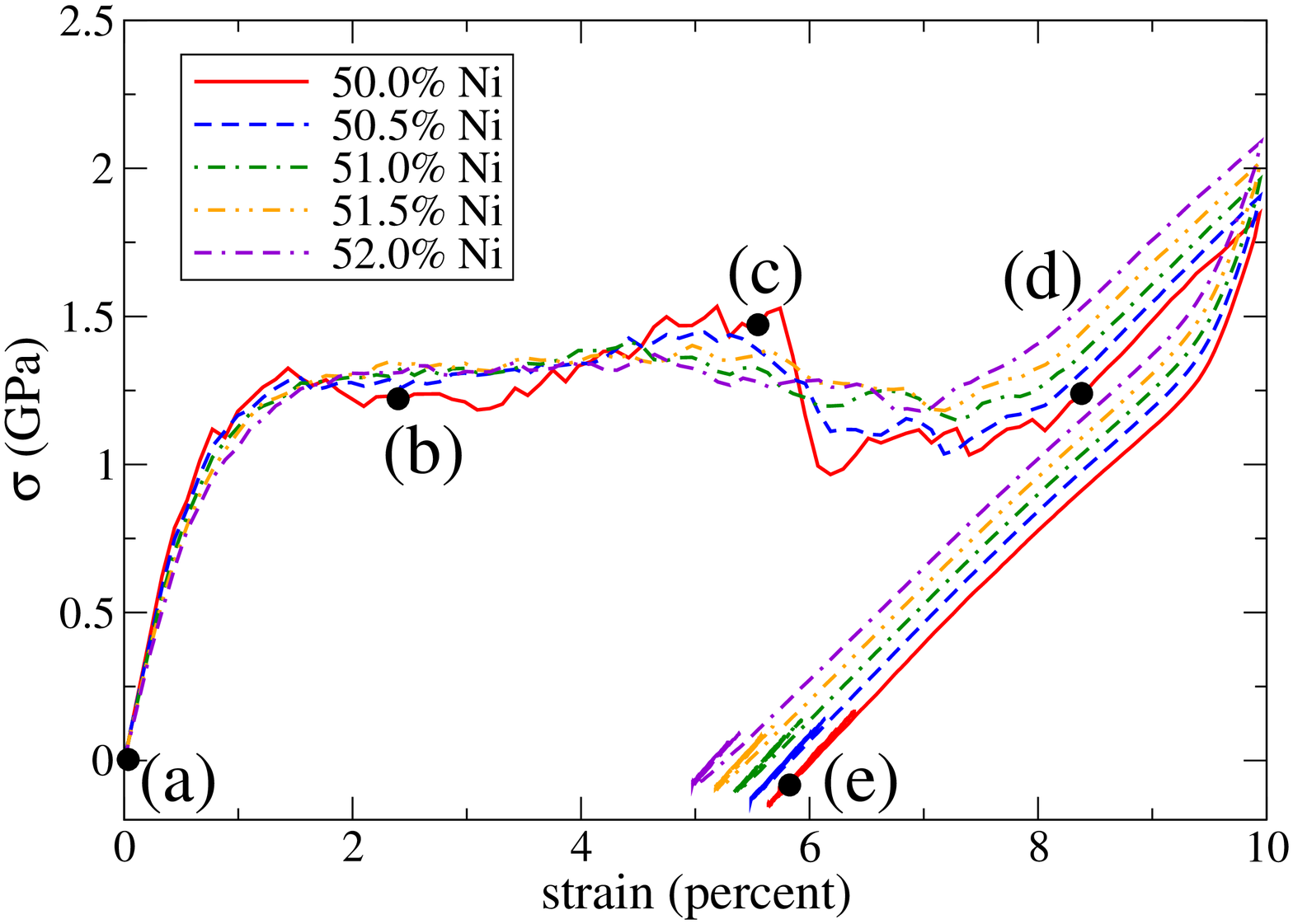}
\caption{Stress-strain curves of the nanoparticle upon tensile loading at $T$\,=\,$1$ K for 5 different Ni concentrations. Marked points (a)-(e) correspond to the visualizations in Fig.\:\ref{f2}.}
\label{f3}
\end{figure}

It is experimentally well known and one of the main reasons for the occurrence of the shape memory effect, that the orientation of martensitic domains changes according to an externally applied load, resulting in a remanent deformation \cite{miyots:89}. In order to observe and study this behavior in the nanoparticle by simulation, a tensile load was applied in x-direction, as indicated by the arrows in Fig.\:\ref{f2}(a). The outermost layer of the y-z-plane in both directions was moved about $10^{-15}$ m at every timestep, leading to a loading rate of 1 m/s. The atoms of these surfaces were prevented from moving back to the equilibrium configuration by setting their velocities and accelerations to zero at every timestep. This lead to a linearly increasing strain of the nanoparticle. The corresponding stress was calculated as the negative internal pressure in the system. Fig.\:\ref{f3} shows the stress-strain curves for five different Ni concentrations at $T$\,=\,1 K. Concentrations varying from 50\% were achieved by choosing a corresponding amount of Ni atoms randomly and replacing them by Ti atoms. The curves represent mean values over five different of those random distributions.
There is a strong stress increase at small strains, corresponding to elastic deformation. The slopes are slightly bigger at concentrations lying closer to the perfect configuration with 50\% Ni, indicating a stiffer elastic behavior in this case. At a critical strain of about 1.8\%, a stress plateau is reached until about 7\% strain. The higher the Ni content, the smoother is the stress in this range, with less abrupt increases and releases. This behavior, as well as the lower stiffnesses at higher Ni contents can be explained by a growing lattice instability, what was figured out to be the reason for the strong decrease of transition temperatures with Ni content in a previous work \cite{mutnie:10}. Figs.\:\ref{f2}(b)-(d) show the development of the microstructure during the loading process. It is seen, that at the critical stress, a reorientation of the B to the A version of the $\overline{\mbox{B19}'}$ structure begins, what is proceeded mainly by twin boundary movement, until a nearly single domain structure remains. At higher strains, the stress increases linear due to elastic deformation again. In the next step of the simulation, the external load was instantaneously removed, leading to a relaxation of the system and vanishing stress. After the relaxation process, which was performed over 50000 timesteps (50 ps), the single domain structure stayed stable [Fig.\:\ref{f2}(e)], accompanied by a remanent strain between about 5.6\% (50\% Ni) and 5\% (52\% Ni).

\subsection{Heating}

\begin{figure}
\centering
\includegraphics[width=0.7\linewidth]{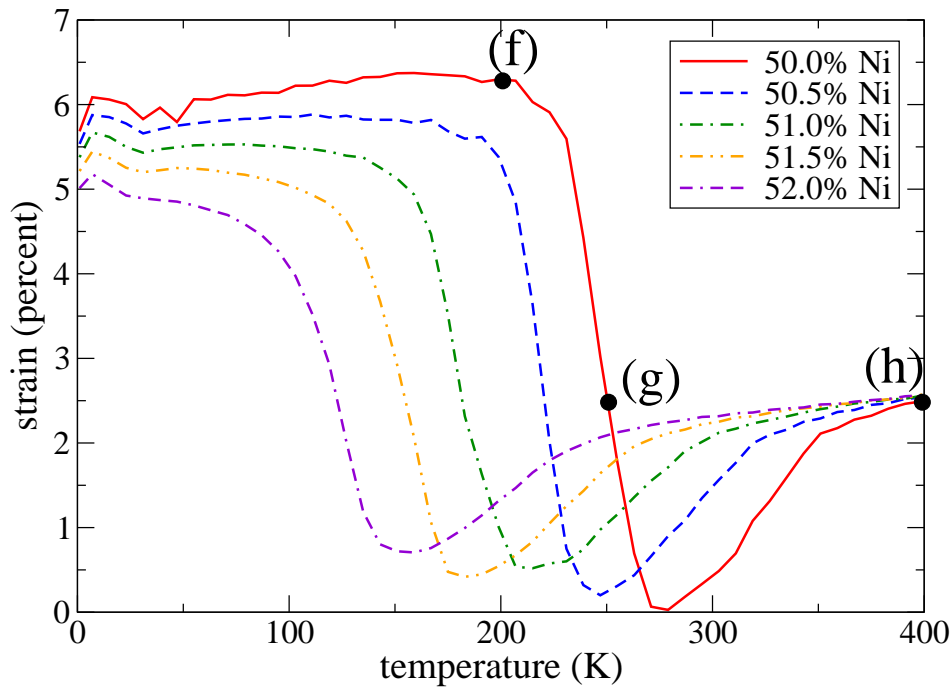}
\caption{Dependence of the strain of the nanoparticle on temperature for 5 different Ni concentrations. Marked points (f)-(h) correspond to the visualizations in Fig.\:\ref{f2}.}
\label{f4}
\end{figure}

As known for macroscopic SMAs, the recovery of the shape, which they had before loading, takes place, if the systems are heated over a certain temperature, at which each martensitic variant adopts the same configuration of a higher symmetric structure (austenite), independent of the initial orientation. In order to analyze, whether and how this effect occurs in the nanoparticle studied in this work, the systems were heated with a rate of 1 K every 1000 timesteps (1 K/ps). Fig.\:\ref{f4} shows the strain behavior during this heating process up to 400 K for the different Ni concentrations. It is common for all systems, that the strain stays almost constant at low temperatures. As depicted in Fig.\:\ref{f2}(f), the martensitic structure containing a single orientation remains stable within this temperature range. At a certain temperature, which depends strongly on concentration, a drop of the strain begins, leading to a nearly complete disappearance. The microstructural evolution during this process is shown in Fig.\:\ref{f2}(g). Instead of a shape recovery due to the transformation to austenite, the initial multi variant configuration is adopted by the system. A subsequent cooling would end up in the starting configuration again. An explanation of this behavior is as follows: both, the single, as well as the multi variant state of the nanoparticle are free energy minima, with the latter being the global minimum, i.e. the ground state of the system. But there is a free energy barrier between these states, corresponding to the energy needed to form twin boundaries and to reorientate parts of the structure. The height of this barrier depends on Ni content, since, as mentioned above, higher Ni concentrations lead to more instable lattice structures, in which twin boundaries can more easily be formed. At a certain temperature, the energy barrier can be overcome by the system, leading to a recovery of the multi variant structure. Further heating beyond this temperature initiates the phase transformation from $\overline{\mbox{B19}'}$ martensite to B2 austenite, ending up in a system with 2.5\% strain, independent of Ni concentration [Fig.\:\ref{f2}(h)]. Cooling below the martensitic transition temperature would trigger the B2$\rightarrow$\,$\overline{\mbox{B19}'}$ phase change, transforming the system back to the initial configuration [Fig.\:\ref{f2}(a)], what is the conventional SME.

\section{Conclusions}

In conclusion, a MD study of the SME of a NiTi model system at the nanoscale is presented. A tensile load was applied to a nanoparticle with a martensitic structure containing variants of different orientation. Due to the load, the boundaries between these twins move and the system reaches a single variant configuration with a remaining non-zero strain after unloading. During a subsequent heating process, a shape recovery different to the conventional mechanism of the SME occurs. It is not the phase transition to austenite which makes the strain vanish, but a transformation back to the ground state configuration with multiple twin orientations, without a change of the crystal structure. The height of the energy barrier, which has to be overcome for this process, decreases significantly, if the perfectly ordered bi-atomic alloy composition is changed slightly. Since the distribution of variant orientations in the martensite depends mainly on surface energy contributions, this kind of SME occurs due to the huge surface-to-volume fraction in nanoscale systems and not in macroscopic SMAs.\\

\begin{acknowledgments}
We gratefully acknowledge the support of the SFB 767.
\end{acknowledgments}

\end{document}